\documentclass[twocolumn,secnumarabic,amssymb,amsmath, nobibnotes, aps, prc, superscriptaddress]{revtex4-2}

\usepackage{mathtools}
\usepackage{graphicx}
\usepackage{color}
\usepackage{hyperref}

\newcommand{\bmat}{\begin{displaymath}}
\newcommand{\emat}{\end{displaymath}}
\newcommand{\beq}{\begin{equation}}
\newcommand{\eeq}{\end{equation}}
\newcommand{\beqa}{\begin{eqnarray}}
\newcommand{\eeqa}{\end{eqnarray}}
\newcommand{\bite}{\begin{itemize}}
\newcommand{\eite}{\end{itemize}}
\newcommand{\bnum}{\begin{enumerate}}
\newcommand{\enum}{\end{enumerate}}
\newcommand{\bdes}{\begin{description}}
\newcommand{\bcen}{\begin{center}}
\newcommand{\ecen}{\end{center}}
\newcommand{\pvec}{\ensuremath{\mathbf{p}}}

\newcommand{\qvec}{\ensuremath{\mathbf{q}}}
\newcommand{\Kvec}{\ensuremath{\mathbf{K}}}
\newcommand{\pt}{\ensuremath{p_{\rm T}}}

\newcommand{\kt}{\ensuremath{K_{\rm T}}}

\newcommand{\qinv}{\ensuremath{Q_{\rm inv}}}

\newcommand{\qout}{\ensuremath{q_{\rm out}}}
\newcommand{\qside}{\ensuremath{q_{\rm side}}}
\newcommand{\qlong}{\ensuremath{q_{\rm long}}}
\newcommand{\qpar}{\ensuremath{q_{||}}}
\newcommand{\qper}{\ensuremath{q_\perp}}
\newcommand{\rout}{\ensuremath{R_{\rm out}}}
\newcommand{\rside}{\ensuremath{R_{\rm side}}}
\newcommand{\rlong}{\ensuremath{R_{\rm long}}}
\newcommand{\rperp}{\ensuremath{R_\perp}}
\newcommand{\minv}{\ensuremath{M_{\rm inv}}}

\newcommand{\alp}{\ensuremath{\alpha}}
\newcommand{\de}{\ensuremath{\Delta E}}
\newcommand{\qsl}{\ensuremath{\sqrt{\qside^2+\qlong^2}}}

\begin{document}

\title{One-, two-, and three-dimensional photon femtoscopy}
\author{D. Mi\'skowiec}
\email[Corresponding author: ]{d.miskowiec@gsi.de}
\affiliation{Research Division and ExtreMe Matter Institute EMMI, 
GSI Helmholtzzentrum für Schwerionenforschung GmbH, Darmstadt, Germany}
\author{K. Reygers}
\affiliation{Physikalisches Institut, Ruprecht-Karls-Universität Heidelberg, Heidelberg, Germany}
\date{\today}
\begin{abstract}
Femtoscopy with photon pairs is a particularly attractive tool for studying high-energy 
nuclear collisions. Proposed and extensively discussed in several influential theory articles, 
it has seen only few applications in experiment because of statistics limitations. 
With the progress of detectors and electronics, only now it is coming within reach. 
In this paper we discuss the choice of kinematic variables for two-photon correlation 
functions. In particular, we argue against $C(\qinv)$ and in favor of $C(\de,\qinv)$. 
\end{abstract}
\maketitle


\section{Introduction}
Femtoscopy, the analysis of momentum correlations between pairs of particles, 
is a powerful and well established technique for assessing the geometry of  
particle-emitting systems created in energetic hadronic and nuclear 
collisions~\cite{Lisa:2005dd}. 
It is most commonly applied to pions, for which the 
correlations are dominated by the Bose-Einstein (BE) enhancement of pairs 
of identical bosons with small relative momenta. 
The width of the BE peak located at zero relative momentum is inversely 
proportional to the width of the distribution of pions of a given momentum 
at the point of their last scattering (kinetic freeze-out). 
This, along with momentum spectra, gives access to the final stage of 
the collision. 

Even more valuable information can be obtained from the analysis of direct photons, 
which  are produced mostly at the beginning of the collision and do not scatter 
subsequently. 
Unfortunately, these photons are buried under a background of 
photons coming from the decays of hadrons, mostly $\pi^0$. The decay photons 
dilute the correlation function, reducing the amplitude of the BE peak by 
a factor of $f^2$, where $f$ is the fraction of direct photons. 
One could argue that this does not spoil the measurement of the geometry 
(width of the peak), and the reduction of the peak height can even be used 
to measure the direct-photon yield, complementing the standard method 
based on the cocktail subtraction~\cite{WA98:2000vxl}. The price for this, however, 
is the necessity to deal with large statistical uncertainties coming from the 
background photons. 

In this situation, the choice of the data representation is particularly important.
In the experimental photon femtoscopy studies published so far the invariant momentum 
difference \qinv\ (see below for definition), well established in identical-hadron 
femtoscopy, has been adopted. Our central claim is that the one-dimensional correlation 
function C(\qinv) is not optimal for photon femtoscopy and should be replaced by a 
two-dimensional analysis.


\section{Photon femtoscopy in theory}
The Bose-Einstein correlations between photons have been the subject of several 
seminal theoretical articles. 
In Ref.~\cite{Neuhauser:1986ija} by Neuhauser, the two-photon correlation functions 
are considered in two dimensions, as a function of their opening angle and the 
energy difference. 
Sinyukov~\cite{Sinyukov:1989xz} discussed a longitudinally expanding source 
of photons and analyzed the correlation function in transverse (to the beam 
axis) and longitudinal momentum difference. 
Srivastava and Kapusta~\cite{Srivastava:1993js} studied the shape of the two-photon 
correlation functions vs the rapidity and transverse momentum differences as 
well as the azimuthal opening angle.  
In the subsequent papers~\cite{Srivastava:1993pt,Srivastava:1994ab} 
they switched to the Bertsch-Pratt~\cite{Bertsch:1989vn, Pratt:1986cc} 
coordinates, which around that time entered the scene and became the de facto 
standard axes in pion femtoscopy. 
The relative momentum here is decomposed into 
\beq
{\bf q} = {\bf p}_2 - {\bf p}_1 = (\qout,\qside,\qlong) \ , 
\eeq
with the “out” axis pointing along the pair 
transverse momentum, the “side” axis perpendicular to it in the transverse 
plane, and the “long” axis along the beam. 
This three-dimensional coordinate system was also used by Bass, Muller, and 
Srivastava~\cite{Bass:2004de}, Frodermann and Heinz~\cite{Frodermann:2009nx}, 
and Alam et al.~\cite{Alam:2003gx}. The latter article also shows 
one-dimensional correlation functions $C(\qinv)$, 
\beq
\qinv = \sqrt{-(p_2^\mu-p_1^\mu)^2} \, ,
\eeq
albeit under the condition that $\qout=\qlong=0$. 

While the two-particle correlation functions of identical and nonidentical 
hadrons exhibit structures coming from quantum statistics and Coulomb and 
strong interactions at small relative velocities (two particles 
``flying together'' in the laboratory), 
for photon pairs the situation is different. The photon velocity in vacuum 
does not depend on the photon energy and thus any two photons emitted close 
to each other in space-time and in the same direction ``fly together'' even 
if they differ in energy. 
In this situation one might wonder if the BE enhancement occurs also for 
such pairs. 
Intuition suggests that this should not be the case, and indeed in all the 
above theory papers the BE peak is restricted to small \qout. 


\section{Photon femtoscopy in experiment}
The experimental data on two-photon Bose-Einstein correlations are scarce. 
The first analysis was performed on data collected in heavy-ion collisions 
at 40--60 MeV per nucleon (i.e.~well below the $\pi^0$ production threshold) 
by TAPS at GANIL~\cite{Ostendorf:1992ah,Marques:1997sw}. 
The correlation function was represented as $C(q,q_0)$, with $q_0$ 
defined as the energy difference $q_0=\de=E_2-E_1$. In order to reduce 
the statistical uncertainties, the $q_0$ dependence was dropped, 
arguing that the energy spectrum of direct photons is exponential, 
hence $q_0$ small, and thus the averaging over $q_0$ does not destroy the 
signal. A significant fraction of photons originated from decays of 
subthreshold $\pi^0$, as clearly seen in the invariant mass spectrum. 
Since for photons $\qinv=\minv$, for the sake of keeping the $\pi^0$ 
peak narrow the authors went one step further and switched from $C(q)$ 
to $C(\qinv)$. 
The same representation was used in a subsequent centrality 
dependent analysis of the same collision system with increased 
statistics~\cite{Piasecki:2006yx}.

At the SPS, the WA98 collaboration published a two-photon $C(\qinv)$ 
measured in central Pb--Pb collisions~\cite{WA98:2003ukc,Peresunko:2004jt,Peressounko:2004ha}. 
The fitted correlation strength (Bose-Einstein peak height) $\lambda$ 
was used to calculate the direct-photon fraction. 

The STAR collaboration at RHIC presented a preliminary $C(\qinv)$ from Au-Au 
collisions in the QM2005 conference~\cite{Das:2005qz}. 
In QM2006, the PHENIX collaboration showed their preliminary results~\cite{Peressounko:2007mp}.  
Also in this case $C(\qinv)$ has been used. 


\section{Abscissa: choice of the kinematic variables}
\label{sec:abscissa}
As shown above, the theory predictions and the experimental measurements 
of two-photon correlation functions so far were performed in terms of 
$C(\qout,\qside,\qlong)$ and $C(\qinv)$, respectively. 
The measurements were plagued by statistical uncertainties resulting from 
small number of events at GANIL and from $\pi^0$ decay background at the 
SPS and above. 
The representation chosen by experimentalists, $C(\qinv)$, keeps the 
$\pi^0$ peak narrow and exposes the flat part of the correlation function 
between the Bose-Einstein peak and $\pi^0$ peak, facilitating the 
normalization of the correlation function~\cite{Marques:1997sw}. 
Let us look more closely at the meaning of \qinv. 
We start from the quantities measured in the experiment, the photon momenta 
(i.e.~energies) $p_1$, $p_2$ and the pair opening angle $\alpha$. 
The invariant momentum difference \qinv\ coincides with the invariant mass 
and is equal to 
\beqa
\qinv &=& \sqrt{2 \, p_1 \, p_2 \, (1-\cos \alp)} \\
      &=& 2 \, \sqrt{p_1 \, p_2} \, \sin \frac{\alp}{2} \ .
\eeqa
In the limits of collinear and back-to-back pairs it becomes
\beq
\qinv = \begin{cases*}
  \sqrt{p_1 p_2} \, \alp \, ,      & for $\alp \rightarrow 0$ , \\
  2 \, \sqrt{p_1 p_2} \, = 2p \, , & for $\alp \rightarrow \pi$ . 
  \end{cases*}
\eeq

As one can see, \qinv\ is not directly sensitive to the energy 
difference between the two photons. 
For pairs of massive identical particles a vanishing \qinv\ implies ${\bf p}_1={\bf p}_2$. 
For photons it does not. 
\qinv\ mostly reflects \qside\ and \qlong\ and is only weakly related to \qout. 
This means that the Bose-Einstein peak, which sits at $\qout=\qside=\qlong=0$, 
in $C(\qinv)$ is diluted by uncorrelated pairs with large \qout. 
This deficiency of $C(\qinv)$ has been noticed 
(Refs.~\cite{WA98:2003ukc,Peresunko:2004jt,Peressounko:2004ha,Peressounko:2007mp,Peressounko:2005ce}; 
see also Eq. (38) in Ref.~\cite{Lisa:2005dd}) and 
the resulting reduction of the peak amplitude $\lambda$ was estimated 
and corrected for by performing a simulation accounting for the detector 
acceptance, the shape of the photon spectrum, and the assumed 
\rout~\cite{Peressounko:2005ce}. 
The direct photon yield published by WA98~\cite{WA98:2003ukc} was 
presented for two cases, $\rout=0$ and $\rout=6$~fm. 
The magnitude of this effect increases with the collision energy (see Appendix). 

Preserving the full height of the BE peak and at the same time keeping 
the $\pi^0$ peak narrow is possible in a two-dimensional analysis 
\beq
C(\de,\qinv) \ . 
\label{de-qinv}
\eeq
The additional axis represents the energy difference between the two 
photons $\de=p_2-p_1$. 
It is easy to show that the length of the vector $(\de,\qinv)$ is equal to $q$ 
\beq
\de^2+\qinv^2 = \qout^2+\qside^2+\qlong^2 \ .
\label{length}
\eeq
The representation (\ref{de-qinv}) is similar to Ref.~\cite{Neuhauser:1986ija} 
except the opening angle is replaced by \qinv, and similar to 
Ref.~\cite{Marques:1997sw} but with $q$ replaced by \qinv. 
The suppression of the BE peak and the systematic uncertainty related to its 
correction are absent in this case. 

\begin{figure}[h]
  \hspace*{-3mm}
  \includegraphics[width=0.5\textwidth]{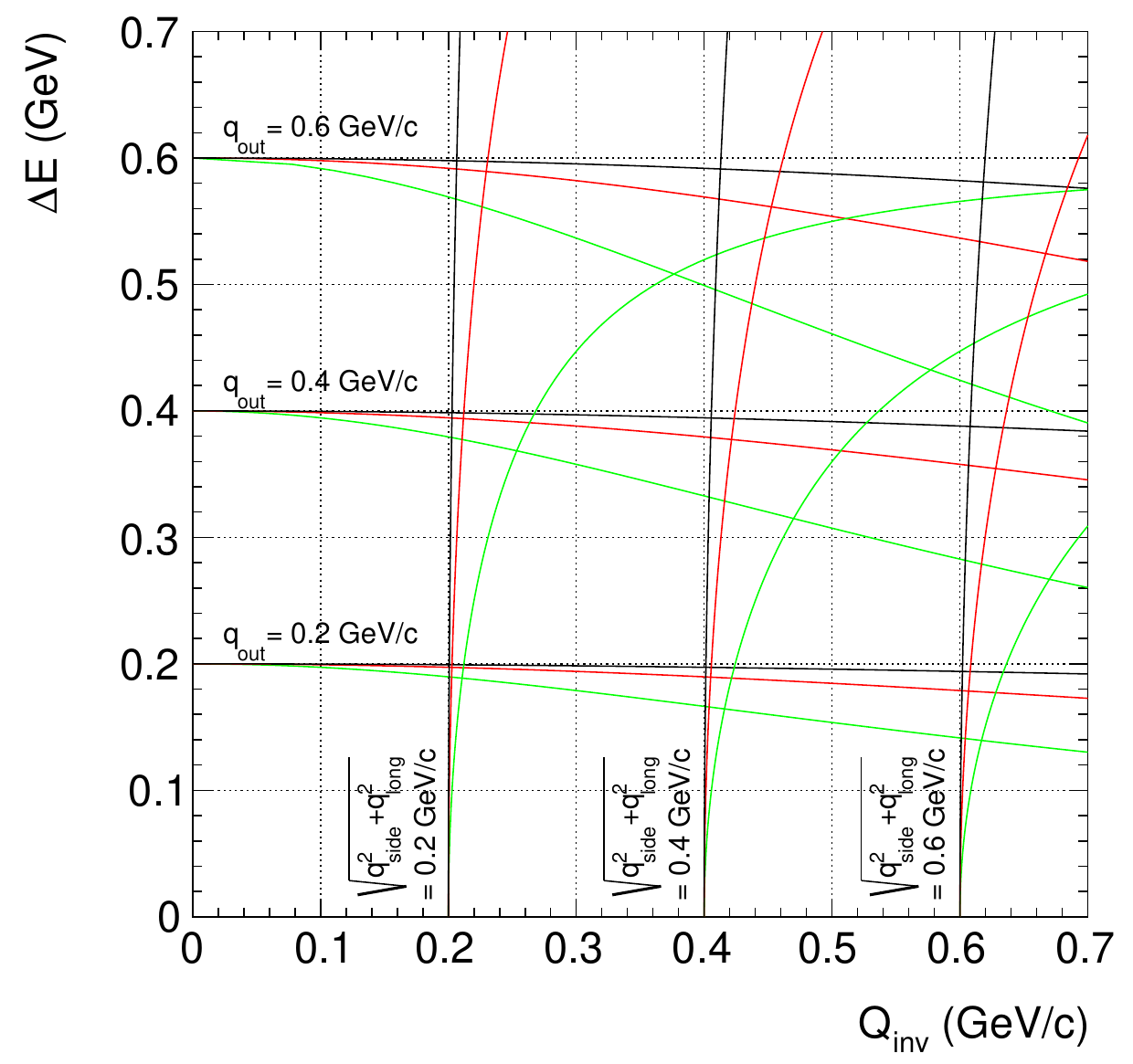}
  \caption{Mapping of $(\qout,\qsl)$ on the $(\de,\qinv)$ plane in LCMS.  
    Lines of constant \qout\ and those of constant $\qsl$ are 
    shown in green, red, and black for the pair momentum $K$ of 0.3, 0.6, and 1.2 GeV/$c$, 
    respectively. In the region below 0.1 GeV/$c$, where the Bose-Einstein peak 
    is expected to be located for direct photons from heavy-ion collisions, the 
    two representations are very close to each other.}
  \label{fig1}
\end{figure}

Our representation (\ref{de-qinv}) is closely related to 
\beq
C(\qout,\qsl) \ .
\label{cosl}
\eeq
In order to demonstrate this, we first note that the components of 
the momentum difference \qvec\ parallel and perpendicular to the pair 
momentum $\Kvec:=(\pvec_1+\pvec_2)/2$ are 
\beq
\qpar = \frac{(\pvec_2-\pvec_1) \cdot (\pvec_1+\pvec_2)}{2 \, K} = \frac{p_2^2-p_1^2}{2 \, K}
\eeq
and
\beq
\qper = \frac{(\pvec_2-\pvec_1) \times (\pvec_1+\pvec_2)}{2 \, K} = \frac{p_1 \, p_2 \, \sin\alp}{K} \ .
\eeq
Secondly, we eliminate $p_1$, $p_2$, $\alpha$ and get the direct relation 
between $(\de,\qinv)$ and $(\qpar,\qper)$: 
\beqa
\qinv^2 &=&\frac{1}{2} \left[q^2-4K^2+\sqrt{(q^2+4K^2)^2-(4K\qpar)^2}\right] \nonumber \\
\de &=& \frac{2K\qpar}{\sqrt{\qinv^2+4K^2}}
\eeqa
with $q^2=\qpar^2+\qper^2=\qout^2+\qside^2+\qlong^2$. 
The inverse transformation reads  
\beqa
\qpar &=& \de \sqrt{1+\left(\frac{\qinv}{2K}\right)^2} \nonumber \\ 
\qper &=& \qinv \sqrt{1-\left(\frac{\de}{2K}\right)^2} \ .
\label{parper}
\eeqa
Third, in the longitudinally comoving coordinate system (LCMS), commonly used in 
femtoscopy~\cite{Lisa:2005dd}, the pair momentum ${\bf K}$ is perpendicular 
to the beam direction, $K=\kt$, and 
\beqa
\qpar &=& \qout \ , \nonumber \\
\qper &=& \qsl \ .
\eeqa
The resulting relation between the $(\de,\qinv)$ coordinates and $(\qout,\qsl)$ 
in LCMS is shown in Fig.~\ref{fig1}. 
In the region below 0.1 GeV/$c$, where the Bose-Einstein peak is expected to 
be located for direct photons from heavy-ion collisions, the two sets of 
variables are very close to each other. 

In order to make a more quantitative statement, let us look at the shape of 
the BE peak in both representations. 
We start from a Gaussian in $C(\qout,\qside,\qlong)$~\cite{Frodermann:2009nx}: 
\beq
C=1+\lambda \exp\left(-\rout^2\qout^2-\rside^2\qside^2-\rlong^2\qlong^2\right) \, . 
\eeq
Its parameters can be directly extracted from a fit to $C(\de,\qinv)$: 
\beqa
C&=&1+\lambda \exp\left(-\rout^2\de^2-\rperp^2\qinv^2\right) \times  \nonumber \\
 &&\times \exp\left[\left(\rperp^2-\rout^2\right)\left(\frac{\de \ \qinv}{2\kt}\right)^2\right] \ . 
\eeqa
Here, \rperp\ is a weighted mean of \rside\ and \rlong, with the weight 
factors depending on the experiment's acceptance and two-photon resolution.  
The exponent in the second line is close to unity for \mbox{$q \ll \kt$} and thus fitting 
the BE peak in $C(\de,\qinv)$ is practically equivalent to fitting it in 
$C(\qout,\qsl)$, as already suggested by Fig.~\ref{fig1}. 

Around midrapidity, 
an additional practical advantage of the representation (\ref{de-qinv}) is that 
the experimental effects related to the determination of the photon momentum 
and the opening angle act nearly separately on the two dimensions of the correlation 
function. The finite opening-angle resolution affects only \qinv, and \de\ is 
mostly affected by the energy resolution. 

Similarly, correlated background may manifest mostly in one of the two dimensions. 
Not only the $\pi^0$ decay but also anisotropic flow and, for calorimeters, 
photon conversion and cluster splitting will result in background pairs with small 
opening angles but not necessarily close in energy. 

At this point we would like to mention an intermediate stage between 
$C(\de,\qinv)$ and $C(\qout,\qsl)$, namely 
\beq
C(\qout,\qinv) \ . 
\eeq
This representation spares the transformation between \de\ and \qout. 
It is, however, less elegant in that only at low \de\ and \qinv\ 
the vector $(\de,0)$ is orthogonal to $(0,\qinv)$ and the length 
is preserved as in Eq.~(\ref{length}). 


\section{Ordinate: statistical uncertainties in various representations}
\label{sec:statistics}
The tendency to reduce the number of dimensions of the correlation function 
is common in experiments encountering large statistical uncertainties. 
Reducing the point-to-point scatter of the correlation function, be it by 
reducing the number of dimensions or by increasing the bin width, has indeed 
a great positive impact on the visual appearance of the correlation. 
However, it does not lead to a more precise fit. 
(If anything, an overly coarse binning can distort the fit result.)  
A properly performed fit -- one based on the maximum likelihood method, assuming 
the Poissonian distribution of the number of counts, and taking its standard 
deviation not from the numerator but from the denominator multiplied by the 
fit function -- can perfectly well deal with low numbers of counts per bin, 
including empty bins~\cite{E877:1996zlh,E802:2002fwe}. 
One-dimensional plots of the correlation function appropriate for visual 
inspection can be constructed as usual by building a ratio between the 
projections of the numerator and denominator. 
The fit function will be visualized by plotting the projection of the denominator 
multiplied by the fit, divided by the projection of the denominator. 
Whenever such projections are made, cuts on the other dimensions will be applied 
such as to maximize the BE signal. 
Following these standard procedures, the additional dimension in $C(\de,\qinv)$, 
compared to $C(\qinv)$, will come at no cost. 


\section{Conclusion}
We advocate using the two-dimensional correlation function $C(\de,\qinv)$ in 
LCMS for analysis of Bose-Einstein correlations of photons in heavy-ion collisions. 
This representation avoids the dilution of the BE peak happening when 
the one-dimensional $C(\qinv)$ is used. At the same time it preserves 
the advantage of the latter, namely it keeps the $\pi^0$ peak narrow. 
The fit precision will not be affected by the additional dimension. 


\begin{acknowledgments}
This work has been supported by the DFG (German Research Foundation) –
Project-ID 273811115 – SFB 1225 ISOQUANT.
\end{acknowledgments}


\appendix*
\section{Dilution of the Bose-Einstein peak in the $C(\qinv)$ representation}
\label{sec:appa}
We estimated the reduction of the peak height $\lambda$ in $C(\qinv)$ by a 
Monte Carlo simulation. We sampled photons from 
\beq
\frac{{\rm d} N}{\pt \, {\rm d}\pt \, {\rm d} y \, {\rm d}\varphi} \propto \frac{1}{\exp(\pt/T)-1} 
\eeq
within $|y|\!<\!1$ and $0\! \le \! \varphi\!<\!2\pi$. 
\begin{figure}[ht]
  \hspace*{-3mm}
  \includegraphics[width=0.45\textwidth]{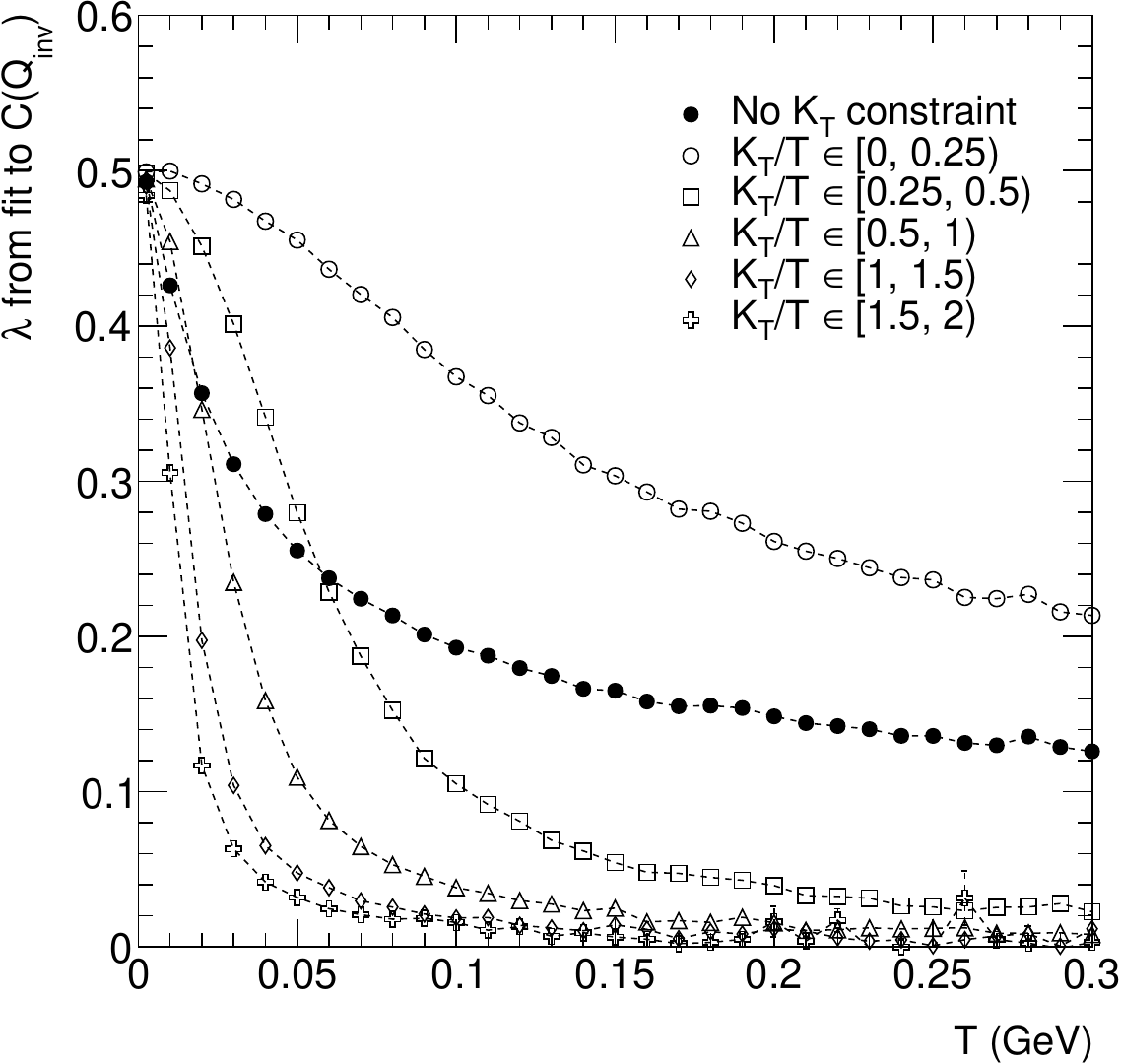}
  \caption{The correlation strength $\lambda$ obtained by fitting a Gaussian to
    $C(\qinv)$ (Monte Carlo simulation).
    The reduction of $\lambda$ is particularly strong at high collision energies
    and high pair momenta.}
  \label{fig2}
\end{figure}
To each photon pair we assigned a weight corresponding to the three-dimensional 
correlation function 
\beq
C=1+0.5 \exp\left(-\rout^2\qout^2-\rside^2\qside^2-\rlong^2\qlong^2\right) 
\eeq
with $\rout \! = \! \rside \! = \! \rlong \! = 6$~fm. 
We histogrammed the pairs with and without the weight factor to obtain, respectively, 
the numerator and denominator of $C(\qinv)$. 
As expected, the correlation strength $\lambda$ (peak height) obtained by fitting a 
Gaussian to $C(\qinv)$ is reduced and the effect is particularly strong at high $T$ 
(high collision energy) and high pair transverse momentum $\kt$ (Fig.~\ref{fig2}). 


\end{document}